\documentstyle[twoside]{article}

\catcode`\@=11
\long\def\@makefntext#1{
\protect\noindent \hbox to 3.2pt {\hskip-.9pt  
$^{{\eightrm\@thefnmark}}$\hfil}#1\hfill}		

\def\thefootnote{\fnsymbol{footnote}}
\def\@makefnmark{\hbox to 0pt{$^{\@thefnmark}$\hss}}	
	
\def\ps@myheadings{\let\@mkboth\@gobbletwo
\def\@oddhead{\hbox{}
\rightmark\hfil\eightrm\thepage}   
\def\@oddfoot{}\def\@evenhead{\eightrm\thepage\hfil
\leftmark\hbox{}}\def\@evenfoot{}
\def\sectionmark##1{}\def\subsectionmark##1{}}

\oddsidemargin=\evensidemargin
\renewcommand{\thefootnote}{\fnsymbol{footnote}}

\newcounter{sectionc}\newcounter{subsectionc}\newcounter{subsubsectionc}
\renewcommand{\section}[1] {\vspace{12pt}\addtocounter{sectionc}{1} 
\setcounter{subsectionc}{0}\setcounter{subsubsectionc}{0}\noindent 
	{\tenbf\thesectionc. #1}\par\vspace{5pt}}
\renewcommand{\subsection}[1] {\vspace{12pt}\addtocounter{subsectionc}{1} 
	\setcounter{subsubsectionc}{0}\noindent 
	{\bf\thesectionc.\thesubsectionc. {\kern1pt \bfit #1}}\par\vspace{5pt}}
\renewcommand{\subsubsection}[1] {\vspace{12pt}\addtocounter{subsubsectionc}{1}
	\noindent{\tenrm\thesectionc.\thesubsectionc.\thesubsubsectionc.
	{\kern1pt \tenit #1}}\par\vspace{5pt}}

\newcounter{appendixc}
\newcounter{subappendixc}[appendixc]
\newcounter{subsubappendixc}[subappendixc]
\renewcommand{\thesubappendixc}{\Alph{appendixc}.\arabic{subappendixc}}
\renewcommand{\thesubsubappendixc}
	{\Alph{appendixc}.\arabic{subappendixc}.\arabic{subsubappendixc}}

\renewcommand{\appendix}[1] {\vspace{12pt}
        \refstepcounter{appendixc}
        \setcounter{figure}{0}
        \setcounter{table}{0}
        \setcounter{lemma}{0}
        \setcounter{theorem}{0}
        \setcounter{corollary}{0}
        \setcounter{definition}{0}
        \setcounter{equation}{0}
        \renewcommand{\thefigure}{\Alph{appendixc}.\arabic{figure}}
        \renewcommand{\thetable}{\Alph{appendixc}.\arabic{table}}
        \renewcommand{\theappendixc}{\Alph{appendixc}}
        \renewcommand{\thelemma}{\Alph{appendixc}.\arabic{lemma}}
        \renewcommand{\thetheorem}{\Alph{appendixc}.\arabic{theorem}}
        \renewcommand{\thedefinition}{\Alph{appendixc}.\arabic{definition}}
        \renewcommand{\thecorollary}{\Alph{appendixc}.\arabic{corollary}}
        \renewcommand{\theequation}{\Alph{appendixc}.\arabic{equation}}
        \noindent{\tenbf Appendix \theappendixc #1}\par\vspace{5pt}}
\newcommand{\subappendix}[1] {\vspace{12pt}
        \refstepcounter{subappendixc}
        \noindent{\bf Appendix \thesubappendixc. {\kern1pt \bfit #1}}
	\par\vspace{5pt}}
\newcommand{\subsubappendix}[1] {\vspace{12pt}
        \refstepcounter{subsubappendixc}
        \noindent{\rm Appendix \thesubsubappendixc. {\kern1pt \tenit #1}}
	\par\vspace{5pt}}

\newcommand{\textlineskip}{\baselineskip=13pt}
\newcommand{\smalllineskip}{\baselineskip=10pt}

\def\eightcirc{
\begin{picture}(0,0)
\put(4.4,1.8){\circle{6.5}}
\end{picture}}
\def\eightcopyright{\eightcirc\kern2.7pt\hbox{\eightrm c}}

\def\abstracts#1#2#3{{
	\centering{\begin{minipage}{4.5in}\baselineskip=10pt\footnotesize
	\parindent=0pt #1\par 
	\parindent=15pt #2\par
	\parindent=15pt #3
	\end{minipage}}\par}} 

\newcommand{\bibit}{\nineit}

\renewenvironment{thebibliography}[1]
	{\frenchspacing
	 \ninerm\baselineskip=11pt
	 \begin{list}{\arabic{enumi}.}
	{\usecounter{enumi}\setlength{\parsep}{0pt}
	 \setlength{\leftmargin 12.7pt}{\rightmargin 0pt} 
	 \setlength{\itemsep}{0pt} \settowidth
	{\labelwidth}{#1.}\sloppy}}{\end{list}}

\newcounter{itemlistc}
\newcounter{romanlistc}
\newcounter{alphlistc}
\newcounter{arabiclistc}

\newcommand{\fcaption}[1]{
        \refstepcounter{figure}
        \setbox\@tempboxa = \hbox{\footnotesize Fig.~\thefigure. #1}
        \ifdim \wd\@tempboxa > 5in
           {\begin{center}
        \parbox{5in}{\footnotesize\smalllineskip Fig.~\thefigure. #1}
            \end{center}}
        \else
             {\begin{center}
             {\footnotesize Fig.~\thefigure. #1}
              \end{center}}
        \fi}

\newcommand{\tcaption}[1]{
        \refstepcounter{table}
        \setbox\@tempboxa = \hbox{\footnotesize Table~\thetable. #1}
        \ifdim \wd\@tempboxa > 5in
           {\begin{center}
        \parbox{5in}{\footnotesize\smalllineskip Table~\thetable. #1}
            \end{center}}
        \else
             {\begin{center}
             {\footnotesize Table~\thetable. #1}
              \end{center}}
        \fi}

\def\@citex[#1]#2{\if@filesw\immediate\write\@auxout
	{\string\citation{#2}}\fi
\def\@citea{}\@cite{\@for\@citeb:=#2\do
	{\@citea\def\@citea{,}\@ifundefined
	{b@\@citeb}{{\bf ?}\@warning
	{Citation `\@citeb' on page \thepage \space undefined}}
	{\csname b@\@citeb\endcsname}}}{#1}}

\newif\if@cghi
\def\cite{\@cghitrue\@ifnextchar [{\@tempswatrue
	\@citex}{\@tempswafalse\@citex[]}}
\def\citelow{\@cghifalse\@ifnextchar [{\@tempswatrue
	\@citex}{\@tempswafalse\@citex[]}}
\def\@cite#1#2{{$\null^{#1}$\if@tempswa\typeout
	{IJCGA warning: optional citation argument 
	ignored: `#2'} \fi}}

\def\pmb#1{\setbox0=\hbox{#1}
	\kern-.025em\copy0\kern-\wd0
	\kern.05em\copy0\kern-\wd0
	\kern-.025em\raise.0433em\box0}

\def\fnt#1#2{\footnotetext{\kern-.3em
	{$^{\mbox{\scriptsize #1}}$}{#2}}}

\def\fpage#1{\begingroup
\voffset=.3in
\thispagestyle{empty}\begin{table}[b]\centerline{\footnotesize #1}
	\end{table}\endgroup}

\font\tenrm=cmr10
\font\tenit=cmti10 
\font\tenbf=cmbx10
\font\bfit=cmbxti10 at 10pt
\font\ninerm=cmr9
\font\nineit=cmti9

\font\eightrm=cmr8

\def\qed{\hbox{${\vcenter{\vbox{			
   \hrule height 0.4pt\hbox{\vrule width 0.4pt height 6pt
   \kern5pt\vrule width 0.4pt}\hrule height 0.4pt}}}$}}

\renewcommand{\thefootnote}{\fnsymbol{footnote}}	

\def\MSbar{\overline{\rm MS}}
\def\mbar{{\overline{m}}}

\def\ie{\hbox{\it i.e.\/}}

\def\bar{\overline}

\def\gsim{{\mathrel{\raise2pt\hbox to 8pt{\raise -5pt\hbox{$\sim$}\hss{$>$}}}}}
\def\rsim{{\mathrel{\raise2pt\hbox to 8pt{\raise -5pt\hbox{$\sim$}\hss{$>$}}}}}
\def\lsim{{\mathrel{\raise2pt\hbox to 8pt{\raise -5pt\hbox{$\sim$}\hss{$<$}}}}}

\begin{document}


\normalsize\textlineskip
\thispagestyle{empty}
\setcounter{page}{1}


\begin{flushright}
{LAUR-00-5684}
\end{flushright}
\vspace*{0.5truein}



\fpage{1}
\centerline{\bf LIGHT QUARK MASSES: A STATUS REPORT AT DPF 2000}

\vspace*{0.37truein}
\centerline{\footnotesize Rajan Gupta\footnote{rajan@lanl.gov}}
\vspace*{0.015truein}
\centerline{\footnotesize\it Theoretical Division, Los Alamos National Laboratory,}
\baselineskip=10pt
\centerline{\footnotesize\it Los, Alamos, New Mexico 87544, USA }
\vspace*{10pt}
\centerline{\footnotesize Kim Maltman\footnote{Permanent address: 
Dept. Mathematics and Statistics, York University, 4700 Keele St.,
Toronto, ON, Canada M3J 1P3.  Work 
supported by the Natural Sciences and Engineering
Research Council of Canada.}}
\vspace*{0.015truein}
\centerline{\footnotesize\it Theory Group, TRIUMF, 4004 Wesbrook Mall, 
Vancouver V6T 2A3 Canada, and}
\baselineskip=10pt
\centerline{\footnotesize\it CSSM, Univ. of Adelaide, Adelaide 5005 Australia}
\vspace*{0.225truein}

\vspace*{0.5truein}

\vspace*{0.21truein}

\abstracts{A summary of the extraction of light quark masses from both
QCD sumrules and lattice QCD simulations is presented. 
The focus is on
providing a careful statement of the potential weaknesses in each
calculation, and on integrating the work of different
collaborations to provide a coherent picture.  }{}{}


%
\section{Introduction}	
\label{sec:intro}
\vspace*{-0.5pt}
\noindent

Significant progress has been made since 1996 
in the determination of the light quark masses
from both lattice QCD (LQCD) and QCD sumrules (SR). The
evolution in time of the ranges quoted by the Particle Data Group\cite{pdg00}
and our best estimates as of DPF2000 are
given below.  Since the scale dependence between $2$ and $1$ GeV is
very large, we quote all results in the $\overline{MS}$ scheme at 
scale $\mu=2$ GeV.  Results at the lower scale $1\ {\rm GeV}$ 
may be obtained via $m_q(1\ {\rm GeV})=1.38\, m_q(2\ {\rm GeV})$ 
where the factor $1.38$ corresponds to 4-loop running, with
$\alpha_s(m_\tau^2)=0.334$\cite{ALEPHalpha}.  In quoting separate
values for $m_u$ and $m_d$ we have employed the ratio
$m_u / m_d = 0.553$, obtained from ChPT.
The basis for these estimates is outlined briefly below.
A more detailed discussion 
will appear in an extended version of this writeup~\cite{USlonger}.
\\
\smallskip\noindent
\begin{center}
\begin{tabular}{ccccc}
         & 1996          & 2000         & Proposed SR    &  Proposed LQCD \\
$ m_u $  & $2-8$    MeV  & $ 1-5 $ MeV  & $2.4-3.8$ MeV  &  $ 2.2-2.7 $ MeV  \\
$ m_d $  & $5-15$   MeV  & $ 3-9 $ MeV  & $4.3-6.9$ MeV  &  $ 3.8-4.9 $ MeV  \\
$ m_s $  & $100-300$ MeV & $ 75-170 $ MeV&$83-130$ MeV  &  $ 78-100  $ MeV  \\
\label{tab:mqcpt}
\end{tabular}
\end{center}
\smallskip

\pagebreak

\textheight=7.8truein
\setcounter{footnote}{0}
\renewcommand{\thefootnote}{\alph{footnote}}

\section{QCD Sum Rule Determinations}
\noindent
We focus on recent ($\geq$1995) extractions
of $m_u+m_d$ and $m_s$, which employ
either Borel (Laplace) transformed sum rules (BSR's) or finite energy
sum rules (FESR's).  
For a typical correlator $\Pi$ with spectral function $\rho$, BSR's
approximate $\rho (s)$ by its OPE version
for $s>s_0$.  Uncertainties in the
the corresponding ``continuum'' spectral contribution 
are small only if
$s_0/M^2$ is significantly $>1$ for Borel masses, $M$,
in the ``stability window''of the analysis.  In contrast, the spectral
ansatz has an unphysical gap if $s_0$ lies 
significantly beyond the squared mass
of the last resonance appearing in the low-$s$ part of the ansatz.
FESR's relate the integral from $0$ to $s_0$ of $\rho (s)w(s)$ (with
$w(s)$ any analytic function)
to the integral of $\Pi (s) w(s)$ over the circle 
$\vert s\vert =s_0$ in the complex $s$-plane.
Using the OPE representation for $\Pi$ in the latter integral is
a potential problem since the OPE is expected to break down 
near the real timelike axis.
A study of the isovector vector (IVV) channel 
using hadronic $\tau$ decay data\cite{kmfesr}
shows that FESR's based on weights $w(s)=s^k$ are {\it not} 
well-satisfied at scales $\sim 2-3$ GeV$^2$ but those based on
``pinched weights'' (those satisfying $w(s_0)=0$), which
we will denote PFESR's,
are, in contrast, very well-satisfied.  

\subsection{The Isovector and Isospinor Pseudoscalar Sum Rules}
\noindent
Since $\partial^\mu A^{ij}_\mu = (m_i+m_j):\bar{q}_i i\gamma_5 q_j$:,
$m_u+m_d$ and $m_s+m_u$ can be determined from
sum rules for the corresponding $ij=ud$ and $us$ correlators.
Recent analyses (BPR\cite{bpr} and its
update, P98\cite{prades98},
for $ij=ud$ and 
DPS\cite{dps} for $ij=us$) employ
the 4-loop $D=0$ OPE expression.  Known $D=4,6$ contributions
are at the few percent level, while other non-perturbative effects
are assumed to be small.
The $\pi$ ($K$) spectral
contributions are known, but those from the
(excited) resonance region are not.  
A modified sum of two Breit-Wigners is used to model the latter,
the overall scale being set by normalizing
the sum of resonance {\it tails} at threshold to values given by 
ChPT.
P98 uses non-PFESR's
($w(s)=1$ and $s$)
and $s_0\sim 2\rightarrow 3.5 \ {\rm GeV}^2$;
DPS use the BSR framework and, 
for $\Lambda_{QCD}^{(3)}=380\ {\rm MeV}$,
$s_0\sim 6\rightarrow 8\ {\rm GeV}^2$.  The results are
$[m_u+m_d](2\ {\rm GeV})=9.8\pm 1.9\ {\rm MeV}$\cite{prades98} and
$m_s(2\ {\rm GeV})=112\pm 18\ {\rm MeV}$\cite{dps}.
PFESR's employing the BPR/P98 ansatze (tuned originally 
using {\it non-PFESR's}) in the same range $s_0\sim 2-3\ {\rm GeV}^2$ 
used by BPR/P98 yield
very poor OPE/spectral integral matches\cite{kmfesr}.
Possible sources of this problem are the resonance normalization prescription
and the use of non-PFESR's at scales where they are not well-satisfied
in the IVV channel.
Potential problems for DPS are the use of an input
assumption about the relative strengths of the 
$K(1460)$ and $K(1830)$ contributions and the existence
of a spectral gap between
$s\sim 4$ and $6\ {\rm GeV}^2$.  
To shift focus to spectral normalization at the
resonance peaks, rather than at threshold, 
an incoherent sum
of Breit-Wigners can be employed, and a PFESR
analysis used to fit both $m_i+m_j$ and
the resonance decay constants\cite{km00bounds}.
The resulting optimized OPE/spectral integral
matches are at the $\sim 1\%$ level or better in the fitting window,
$2.8\rightarrow 3.6\ {\rm GeV}^2$, and correspond to
$[m_u+m_d](2\ {\rm GeV})=9.9\pm 1.0 \ {\rm MeV}$,
$m_s(2\ {\rm GeV})=116\pm 5\ {\rm MeV}$.
The above results neglect direct instantons, and
other non-perturbative effects not present in the OPE.
Two examples show that non-trivial
uncertainties are associated with this neglect.  First,
a phenomenologically-determined effective
tachyonic gluon mass-squared, meant to represent additional
short-distance non-perturbative contributions, 
lowers the P98 value by $5.6\%$\cite{cnz99}.
Second, including instanton liquid model 
(ILM)\cite{ilm} estimates of direct instanton effects,
the PFESR extraction, $[m_u+m_d](2\ {\rm GeV})=9.9\ {\rm MeV}$ 
is reduced to $7.8\ {\rm MeV}$\cite{km00bounds}. 

\subsection{The Light-Strange Scalar Sum Rule}
\noindent
Since $\partial^\mu V^{su}_\mu = i(m_s-m_u):\bar{s}u:$,
$m_s-m_u\simeq m_s$ can be
determined from sum rules for the strange scalar channel.
The low-$s$ part of $\rho (s)$ in this case
can be determined indirectly, {\it with certain additional theoretical
assumptions}, from the Omnes representation of the timelike $K\pi$ scalar
formfactor, $d_{K\pi}$\cite{jm}.
JM\cite{jm}, CDPS\cite{cdps}, and 
CPS\cite{cps} employed this construction only to fix $d_{K\pi}$ 
at threshold; the tail of a sum-of-Breit-Wigners was then normalized 
to this value, as in the BPR/P98 analyses.
A purely resonant phase for $d_{K\pi}$, however,
yields a slope, $d^\prime_{K\pi}(s=0)$, incompatible with 
ChPT\cite{cfnp} and a poor OPE/spectral integral
match\cite{kmfesr}; both problems are removed if one includes background
contributions to the phase near threshold and uses the Omnes
representation for all $s$\cite{cfnp,kmss}.
More recent analyses (J98\cite{jamin98}, M99\cite{kmss}) employ the
CFNP version of $\rho (s)$.  CFNP and J98 are BSR analyses differing
only in the value of $s_0$, while M99 is a PFESR analysis.  CFNP
has a good stability plateau for $m_s$, but at the cost of a gap in
$\rho$.  J98 has no spectral gap, but also no stability plateau, and
also potentially non-trivial continuum contributions.  The M99 PFESR's
show an excellent OPE/spectral integral match.  The results,
$m_s(2\ {\rm GeV})=91\rightarrow 116\ {\rm MeV}$ for CFNP, $116\pm 22\
{\rm MeV}$ for J98, and $115\pm 8\ {\rm MeV}$ for M99, are all
compatible.  {\it Additional} errors associated with the assumptions
made in writing down the Omnes representation are at least $\sim 10\
{\rm MeV}$ in magnitude.

\subsection{Sum Rules Based on Electromagnetic Hadroproduction Data}
\noindent
Sum rules based on electromagnetic (EM) hadroproduction 
data\cite{narison95,km3388,kmcw99,narison99} have significant
uncertainties associated with either (1) isospin-breaking
corrections\cite{km3388,kmcw99} (for the flavor $33$-$88$ vector
current sum rule\cite{narison95}) or
(2) deviations from ideal mixing/Zweig rule violations\cite{km00}
(for the flavor $33$-$ss$ vector current
sum rule\cite{narison99}).  We, therefore, consider
these sumrules uncompetitive with
those based on hadronic $\tau$ decay and defer further discussion
to the extended version of this write-up.

\subsection{Sum Rules Based on Hadronic Tau Decay Data}
\noindent
The ratio of the hadronic $\tau$ decay rate through the flavor $f=ij=ud,us$ 
vector (V) or axial vector (A) current to the 
$\tau\rightarrow \nu_\tau e\bar{\nu}_e$ rate,
$R^{V/A;ij}_\tau $,
can be written as a sum of appropriately-weighted integrals
of the $J=0,1$ parts of
the corresponding hadronic spectral function.
Since the $ij=ud$ and $us$ correlators are identical
in the $SU(3)_F$ limit, rescaled differences such as
$R_{\tau ;ud}/\vert V_{ud}\vert^2 -
R_{\tau ;us}/\vert V_{us}\vert^2\equiv \Delta R_\tau$
(with $V_{ij}$ the $ij$ CKM matrix element and
$R_{\tau ;ij}\equiv R_\tau^{V;ij}+R_\tau^{A;ij}$) 
vanish in this limit.
Experimental data allows access to the $ud$, $us$ spectral
distributions, and hence to integrals of such 
correlator differences, whose OPE representations, for large enough
$s_0$, should be dominated
by the $D=2$ term, proportional
(neglecting $m_{u,d}$) to $m_s^2$.
The kinematic weights appearing 
on both sides of the $\Delta R_\tau$
sum rule may be supplemented by a
factor $(1-s/m_\tau^2)^k(s/m_\tau^2)^l$
without necessitating a $J=0$/$J=1$ experimental decomposition;
the analysis
is then said to employ the ``$(k,l)$ spectral weight''.  

The $D=2$ OPE representations of the $J=0+1$ and $J=0$ parts of 
the $ud$-$us$ correlator are known
to 3-loop ($O(\alpha_s^2)$)
and 4-loop ($O(\alpha_s^3)$) order, respectively.
The $ud$ spectral distributions are very accurately determined\cite{ALEPHalpha}
while the $V+A$ sum for the $us$ case is known
with errors of $\sim 6-8\%$ in the $K^*$ region 
and $\sim 20-30\%$
above the $K^*$\cite{ALEPHstrange}.  All analyses
employ the ALEPH $ud$ and $us$ data.  
The CDH98\cite{cdh}, MT98\cite{kmtauprob}, and CKP98\cite{ckp98}
analyses are based on the preliminary result,
$R_{\tau ,us}=0.155\pm .006$;
ALEPH99\cite{ALEPHstrange}, PP99\cite{pp99}, KKP00\cite{kkp002}
and KM00\cite{km00} on the
1999 published version, $0.1610\pm .0066$\cite{ALEPHstrange}; and
DHPPC00\cite{dhppc00} on a recent update, $0.1630\pm .0057$\cite{dhppc00}. 
Increasing $R_{\tau ,us}$
decreases $R_{\tau ;ud}$, which is obtained via
$R_{\tau ;ud}=R_{\tau ;had}-R_{\tau ;us}=[(1-B_e-B_\mu )/B_e]-R_{\tau ;us}$.
Small increases in $B_e$ and $B_\mu$\cite{pdg00}
have also lowered $R_{\tau ;had}$ since the earlier
ALEPH analyses. The high degree of
cancellation between $ud$ and $us$ spectral integrals
(typically to better than $10\%$) means
the result for $m_s$ is {\it very} sensitive to exactly
which values of $R_{\tau ;ud}$, $R_{\tau ;us}$ have been used.  
For the same reason, $m_s$ is also sensitive
to small variations in the input
values of $f_K$ and $\vert V_{us}\vert$ ($\vert V_{us}\vert^2$, e.g.,
differs by $2.6\%$ depending on whether one uses the value
from $K_{e3}$ or that from 
3-family CKM unitarity, combined
with $\vert V_{ud}\vert$ as extracted
from $0^+\rightarrow 0^+$ nuclear decays; the difference is non-negligible
on the scale of the $<10\%$ $ud$-$us$ residual).  
Different prescriptions for truncating the $D=2$ OPE series
and estimating the truncation error also are present
in the literature.  In Table 1 we display the impact of
converting existing results so as to correspond, where possible, to
(1) the common truncation and error estimate scheme
employed by DHPPC00 and (2) $R_{\tau ;us}=.163$.  A subtlety
is involved in the second of these conversions 
for non-inclusive analyses and/or those employing $s_0\not= m_\tau^2$;
see the extended version of this write-up for details.
Since the $O([\alpha^{eff}_s]^3)$ estimate in the KKP00 framework 
(which employs an effective running coupling and running masses)
is not available to us, the 
KKP00 results have been converted only to reflect the
new values of the $ud$ and $us$ integrals.
Two converted versions are given in each case, one corresponding to
the PDG2000 unitarity-constrained CKM set, 
$\vert V_{ud}\vert =.9479$, $\vert V_{us}\vert =.2225$ (CKMU),
the other to the PDG2000 non-unitarity-constrained set,
$\vert V_{ud}\vert =.9735$, $\vert V_{us}\vert =.2196$ (CKMN).
Results from CDH98\cite{cdh} and MT98\cite{kmtauprob} 
have been omitted since the
former employed an older, erroneous 
expression for the integrated $D=2$, $J=0$ OPE contribution and
the latter an assumption about the relation 
between the $ud$ and $us$
$J=0$ spectral integrals now known to be invalid\cite{km00}.
Also not quoted are the PP99 and ALEPH99 results,
which have been superceded by the joint DHPPC00 update.
The first error quoted is experimental,
the second theoretical.
The reader should bear in mind that the
integrated $D=2$ OPE series for the $J=0$
contribution to the $ud$-$us$ $\tau$ decay sum rules
converges very badly\cite{kmtauprob,pp98,kkp001} (for $s_0=m_\tau^2$ the
ALEPH99 version, for example, is
$\sim 1+0.78+0.78+0.90+\cdots$\cite{ALEPHstrange}).
This creates potentially large theoretical errors 
for those (``inclusive'') analyses
which retain
both the $J=0+1$ and $J=0$ contributions
(ALEPH99, CKP98, PP99, KKP00 and DHPPC00).
While the identification,
and hence subtraction, of the $\pi$ and $K$ pole $J=0$ contributions
is unambiguous, at present
no $J=0$/$J=1$ separation exists for the experimental data in
the excited resonance region.  
Analyses which subtract the $J=0$ contributions (ALEPH99, KM00),
and work with the theoretically
much better behaved $J=0+1$ contribution (the ``0+1'' approach),
thus have instead $J=0$ subtraction uncertainties.

Our summary of estimates based on hadronic $\tau$ decay is given in
Table~\ref{tab:taudecay} along with the analysis type.  Apart from
KM00, these results neglect possible $D>6$ contributions.  Because the
polynomial coefficients grow rapidly with $k$ for the $(k,0)$ spectral
weights, this neglect becomes less safe as $k$ increases.  An extended
discussion of the issues alluded to above, as well
as details on the conversion of the individual results to the values 
quoted, will be given in the longer
writeup~\cite{USlonger}.

\begin{table}[htbp]
\tcaption{Impact of conversion to common input for the
hadronic $\tau$ decay extractions of $m_s$.  Results
for $m_s(2\ {\rm GeV})$ are in MeV, and correspond
to $B_e=.1783$, $B_\mu=.1737$, $R_{\tau ;us}=.163$.
All except KKP00
also correspond to the DHPPC00 truncation
prescription, and to $\alpha_s(m_\tau^2)=.334$.
Experimental and theoretical errors have been combined in quadrature.}
\centerline{\footnotesize\smalllineskip
\begin{tabular}{l l l l l}\\
\hline
&&&$m_s(2\ {\rm GeV})$ (MeV)& \\
\hline
{Reference} &{Analysis Type} &{Original} &CKMU input&CKMN input\\
\hline
CKP98&inclusive, $(0,0)$&$145\pm 36$
&$116\pm 31$&$99\pm 34$\\
KKP00&inclusive, $(0,0)$&$125\pm 28$&$120\pm 28$
&$106\pm 32$\\
KM00&$(0+1)$, $w_{20}$\cite{km00}&$115\pm 17$&$110\pm 16$
&$100\pm 18$\\
DHPPC00&inclusive, $(0,0)$&$126\pm 31$&$124\pm 32$
&$106\pm 37$\\
DHPPC00&inclusive, $(1,0)$&$115\pm 19$&$113\pm 19$
&$102\pm 21$\\
DHPPC00&inclusive, $(2,0)$&$100\pm 21$&$99\pm 21$
&$91\pm 21$\\
\hline\\
\end{tabular}}
\label{tab:taudecay}
\end{table}

\subsection{Summary}
We consider results for $m_s$
based on sum rules involving hadronic $\tau$ decay data both most
reliable and to have the greatest potential for improvement
(through improved $us$ spectral data which should be provided
by the B factories).  Results from other analyses are
compatible, but with less controlled uncertainties.
The agreement between all approaches, however, suggests that
theoretical uncertainties are under reasonable control.
Experimental errors in the different $\tau$
analyses are necessarily
strongly correlated, and hence cannot be averaged.  Consistency
of the different versions does, however, allow us to average
the central values, and quote as our error the smallest of
the errors shown (which corresponds to the least strong 
$ud$-$us$ cancellation).  With this prescription,
we obtain for our final sum rule determination
$m_s(2\ {\rm GeV})=114\pm 16\ {\rm MeV}$ for the
CKMU input set and $m_s(2\ {\rm GeV})= 101\pm 18\ {\rm MeV}$
for the CKMN input set. As mentioned above, theoretical uncertainties 
in $m_u + m_d$ are larger. 
Using the ChPT ratio $2m_s/[m_u+m_d]=24.4\pm 1.5$, the CKMU and
CKMN $m_s$ results
correspond to $[m_u+m_d](2\ {\rm GeV})=9.3\pm 1.4$ and
$8.3\pm 1.6$ MeV, respectively.
For comparison, 
the P98 estimate is either
$9.8$ MeV or, if one uses the P98 $1$ GeV value together
with our scaling factor, $12.8/1.38 = 9.3$ MeV,
while the PFESR/ILM estimate is $7.8$ MeV.
For our final summary we choose the conservative combined
ranges $6.7 - 10.7$ MeV for 
$m_u + m_d$ and $83 - 130$ MeV for $m_s$.  These, as we show next, are in 
excellent accord with LQCD estimates.




\section{Lattice QCD: Quenched Results}
\label{sec:quench}
LQCD simulations provide estimates for two quantities $\mbar \equiv
(m_u+m_d)/2$, and $m_s$.  $m_u$ and $m_d$ are not obtained separately
since current simulations do not include EM effects. 
Details of how quark masses are extracted from lattice calculations
are given in~\cite{Mqrev97}${}^-$\cite{MqCPPACS00} and will not be
repeated here.  We start with a summary of the state-of-the-art
results given in Table~\ref{t:mq} and discuss current shortcomings and
future prospects.  This table also provides information on (i) the
lattice action used in the simulations, (ii) whether extrapolation to
the continuum limit was done, and (iii) the states used to fix the
quark masses and the lattice spacing $a$.  The errors quoted include
statistical and those due to renormalization constants and chiral and
continuum extrapolations. Comparing data from different collaborations
suggests that the quoted statistical errors are realistic and much
larger than finite volume corrections. The latter three systematic
errors will be discussed later.

\begin{table}[ht]
\tcaption{Recent results for quark masses. $O(a)$ SW stands for
non-perturbative $O(a)$ improved Sheikhholeslami-Wohlert (SW) fermion
action, Iwasaki for an improved gauge action, and DWF for Domain Wall
Fermions. In the last column we give the quantity used to set the
lattice spacing $a$ and $(a\to 0)$ indicates that results were extrapolated
to the continuum limit. $r_0$ is defined in terms of the force at
$0.5$ fermi and $a(r_0) \approx a(M_\rho)$.}

\begin{center}
\noindent
\setlength{\tabcolsep}{4.2pt}
\begin{tabular}{|l|l|c|c|c|c|}
\hline
                         &  Action    & $\bar m$        & $m_s(M_K)$  & $m_s(M_\phi)$& scale $1/a$    \\
                         &            &                 &             &              &  (GeV)         \\
\hline
Summary                  &            & $3.8(1)(3)$     & $99(3)(8)$  & $111(7)(20)$ & $M_\rho$       \\
(1997)\cite{Mqrev97}     &            &                 &             &              & ($a\to 0$)     \\
APE                      & O(a) SW    & $4.5(4)$        & $111(12)$   &              & $M_{K^*}$      \\
(1998)\cite{MqAPE98}     &            &                 &             &              & $a^{-1} \approx 2.7$     \\
APE                      & O(a) SW    & $4.8(5)$        & $111(9) $   &              & $M_{K^*}$      \\
(1999)\cite{MqAPE99}     &            &                 &             &              & $a^{-1} \approx 2.7$     \\
JLQCD                    &  Staggered & $4.23(29)$      & $106(7)$    & $129(12)$    & $M_\rho$       \\
(1999)\cite{MqJLQCD99}   &            &                 &             &              & ($a\to 0$)     \\
CPPACS                   &  Wilson    & $4.55(18)$      & $115(2)$    & $143(6)$     & $M_\rho$       \\
(1999)\cite{MqCPPACS99}  &            &                 &             &              & ($a\to 0$)     \\
CP-PACS                  & Iwasaki+SW & $4.4(2)$        & $110(4)$    & $132(6)$     & $M_\rho$       \\
(2000)\cite{MqCPPACS00}   &            &                 &             &              & ($a\to 0$)     \\
ALPHA-UKQCD              & O(a) SW    &                 & $ 97(4)$    &              & $f_K$          \\
(1999) \cite{MqALPHA99}  &            &                 &             &              & ($a\to 0$)     \\
QCDSF                    & O(a) SW    & $4.4(2)$        & $105(4)$    &              & $r_0 $         \\
(1999)\cite{MqQCDSF99}   &            &                 &             &              & ($a\to 0$)     \\
QCDSF                    & Wilson     & $3.8(6)$        & $87(15)$    &              & $r_0 $         \\
(1999)\cite{MqQCDSF99}   &            &                 &             &              & ($a\to 0$)     \\
RIKEN-BNL                & DWF        &                 & $105(6)(21)$& $127(6)(25)$ & $M_\rho$       \\
(2000)\cite{MqBNL00}     &            &                 &             &              & $a^{-1}\approx 2.1$    \\
\hline
\hline
CP-PACS                  & $n_f=2$    & $3.44^{+0.14}_{-0.22}$ & $88^{+4}_{-6}$ & $90^{+5}_{-11}$ & $M_\rho$  \\
(2000)\cite{MqCPPACS00}  & Iwasaki+SW &                 &             &              & ($a\to 0$)     \\
QCDSF-UKQCD              & $n_f=2$    & $3.5(2)$        & $90(5)$     &              & $r_0$          \\
(2000)\cite{MqQCDSFUKQCD00}& O(a) SW    &                 &             &              & $[1.9-2.2]$  \\
\hline
\end{tabular}
\label{t:mq}
\end{center}
\end{table}
%

At first sight, these estimates suggest a lack of consistency between
different lattice calculations. We will attempt to argue that this is
not so.  The key reason is that it is only for QCD with three
dynamical light flavors (the physical theory) that we expect all
simulations to give the same results once the extrapolation to the
continuum limit ($a=0$ to remove discretization errors) has been
carried out. So, before analysing the relative merits of the different
calculations one has to address the question of concern to all: how
valid is the quenched approximation, especially since the resulting
theory is not even hermitian? From a practical point of view quenching
introduces two main limitations. First, the spectrum of the quenched
theory will not coincide with the experimentally observed one. The
second limitation, discussed in~\cite{BGpql,SRSpql}, and known as the problem of
quenched chiral logs, is more subtle.  In a nutshell the problem is
that the quenched $\eta'$ persists as a Goldstone boson in the chiral
limit and its propagator has a single and double pole, consequently
the chiral expansion of pseudoscalar meson masses, decay constants,
and quark condensates develop enhanced logarithms that are singular in
the chiral limit~\cite{BGpql,SRSpql}. The first limitation implies that
there will be no consistent set of quark masses 
which reproduces the observed spectrum in the quenched 
approximation. The second is relevant when
extrapolating data to the physical $u$ and $d$ quark masses, where the
effects of the artifacts (enhanced chiral logs) becomes
significant. So, in fact, the validity of quenched calculations will
be judged {\it a posteriori} by comparison with $(2+1)$ flavor calculations. 

Quenched LQCD simulations proceed with the assumption that the stable
particle masses and decay constants are affected at roughly the $ 10\%$
level, $i.e.$ the deviations are ``small''. Within this assumption,
the quenched results can be presented in one of two ways: (${\cal
P}_1$) analyze the data in all possible ways and take the variation
in the estimates as a measure of the quenching uncertainty,
or (${\cal P}_2$) find combinations of quantities for which quenching
effects are expected to be small, and use only these to extract a
quenched number (which is then presumably ``closest'' to the real world).  
The results in
Table~\ref{t:mq} represent both approaches; for example, calculations
by JLQCD and CP-PACS are closer to ${\cal P}_1$, while
ALPHA-UKQCD follow the ${\cal P}_2$ philosophy.

We start with three concensus statements regarding the quenched data
in Table~\ref{t:mq}. First, for a given gauge and fermion action, the
raw lattice data, and the masses of hadrons as a function of the bare quark
masses, are consistent between the different calculations. Second,
there is $\sim 10\%$ uncertainty in the values of quark masses
depending on what physical quantity ($M_\rho$, $M_{K*}$, $f_K$,
$f_\pi$, or $r_0$) is used to set the lattice scale. (In this case,
for example, proponents of ${\cal P}_2$ would argue that $M_\rho$
should not be used to set $a$ because of its large width and similarly
large shift in the mass between the unmixed and physical state.)
Third, there is $\sim 20\%$ uncertainty in $m_s$ depending on whether
$M_K$ or $M_{K*}$ or $M_\phi$ is used to fix it ($M_{K*}$ and $M_\phi$
give consistent results). Furthermore, as expected, neither $m_s(M_K)$
nor $m_s(M_\phi)$ yield the observed splittings in the baryon
decuplet~\cite{hmLANL97,MqCPPACS99}. Our view, in summary, is that
these variations, with the states used to fix $a$ and the quark
masses, provide a very useful handle for evaluating the ``reality'' of
coming simulations with dynamical quarks: these differences should
vanish as the input dynamical quark masses are tuned closer and closer
to their physical values.

Three other factors give rise to some of the differences
observed in Table~\ref{t:mq}. These are (i) chiral extrapolations (ii)
continuum extrapolations, (iii) the renormalization factor connecting
lattice quark masses to those in the $\MSbar$ scheme at $\mu = 2$
GeV. A brief discussion of these is as follows.

{\it Chiral Extrapolations}: Simulations with physical masses for
$u$ and $d$ quarks are computationally too expensive and beyond the
capability of today's computers. Consequently one simulates the theory
over a range of quark masses, typically $2m_s \rightarrow m_s/3$, and
extrapolates to $\overline{m}$ using predictions of (quenched,
or partially quenched for $n_f=2$) chiral perturbation theory.

{\it Continuum extrapolations}: The lattice theory at scale $a$ has
discretization errors of $O(\alpha_s^i a^n)$, where $i$, $n$ depends on
the order of improvement of the lattice action and operators.  These
errors are removed by extrapolating results calculated over a range of
scales, typically $2 \leq 1/a \leq 4$ GeV, to $a=0$.

{\it Renormalization Constants}: These connect the lattice quark
masses to either the renormalization group invariant
mass~\cite{MqALPHA99}, or those in a continuum renormalization scheme
like $\MSbar$ at some fixed scale. These factors have been estimated
using perturbation theory (CP-PACS), semi non-perturbatively (APE), and fully
non-perturbatively (ALPHA-UKQCD); the latter being the most accurate.

One way to improve the reliability of the chiral and continuum
extrapolations is to use quantities (or their ratios) that have the
least ambiguity and dependence on $m$ or $a$ (${\cal P}_2$
approach). For example, while the statistical signal for extracting
$M_\rho$ is good, using it to set the lattice spacing can lead to a
$\sim 10\%$ uncertainty since the mass of the stable $\rho$ calculated 
in lattice simulations could differ by $\sim 100$ MeV from the physical value. 
It is therefore desirable to use ratios with the best statistical signal,
and the smallest variation with the quark masses or $a$ (see
Ref.~\cite{MqALPHA99} for discussion).


It is not straightforward to assess the absolute magnitude of these
three errors, independent of the analyses.  The reason is that
they are correlated and depend on the details of how each analysis was
done.
For example, consider doing a continuum extrapolation with simulations at
$1/a = 2$, $3$, and $4$ GeV, scales typical of the best
calculations. If the relative (systematic) error at the three points,
due to say the chiral extrapolation, is not the same, then the bias
can get magnified by the continuum extrapolation.  From a survey of
the various calculations, our estimate of the range of uncertainty in
$m_s$ from these three effects varies between $2 - 15$ MeV, with the
lower value coming from the ALPHA-UKQCD result which includes a fully
non-perturbative calculation of the renormalization constant, and
considers a kaon composed of two quarks of mass $\approx m_s/2$ rather
than $m_s$ and $m_u$. The latter simplification avoids the need for
chiral extrapolation in $m_u$ and is justified on the basis that the
chiral fit, $M_\pi^2 $ versus $m$, shows a linear behavior in the
range $2 m_s - m_s/2$.

Quenched calculations have significantly improved our understanding
of, and ability to control, these three sources of systematic
errors.  For example, (i) use of improved actions has lead to a
significant reduction in discretization errors (with even just $O(a)$
improvement, the residual error now is only $5-10\%$ at $1/a \sim 2$
GeV); (ii) development of reliable methods for calculating the
renormalization constants non-perturbatively has reduced the
associated uncertaintly to $< 2\%$~\cite{MqALPHA99}; and (iii) partially
quenched ChPT~\cite{SRSpql00} has provided us with
a much improved understanding of how to extract physical results from heavier
quarks (in the range $2m_s - m_s/4$), especially in the case of future $2+1$
dynamical flavor simulations. To summarize, our bottom line, based
on quenched simulations, is

$\bullet$ The technology for all aspects of the calculation has been developed.

$\bullet$ We have a very good understanding of statistical 
and systematic errors.

$\bullet$ The current best estimates are: $90 \leq m_s \leq 140 $ MeV
and $3.8 \leq \overline{m} \leq 4.8$ MeV.

\section{Lattice QCD: Results with $2$ dynamical flavors}
\label{sec:dynamical}

Until very recently, the least controlled systematic uncertainty, as
discussed above, was that due to the quenched approximation, which was
forced on us by limitations in computer resources. As of this year
a number of collaborations (CP-PACS, JLQCD, APE, QCDSF-UKQCD, MILC,
SESAM) have begun to report results from simulations with two
dynamical flavors.  These were reviewed at Lattice 2000 by
Lubicz~\cite{MqLUBICZ00} and we discuss them briefly later. For the
most part, we restrict our attention to the recent published results
from the CP-PACS collaboration~\cite{MqCPPACS00} (see Table~\ref{t:mq})
as these results are of quality similar to the quenched calculations
in terms of statistics, lattice sizes, and range of $a$ values
explored. 


The CP-PACS results are remarkable in that they show amazing
consistency between $m_s(M_K)$ and $m_s(M_\phi)$, in contrast to the
$\sim 20\%$ difference seen in the quenched theory.  Also, compared to
the quenched results of the ALPHA-UKQCD collaboration 
(who analyze data under
the ${\cal P}_2$ philosophy), these are $\approx 10\%$ lower; the
``expected'' size of quenching errors. 

There, however, remain issues that need be be better understood. First,
comparing CP-PACS quenched and unquenched calculations with the same
gauge and fermion action and with similar analysis methods, the unquenching
effect is much $>10\%$.  For example, $m_s(M_K) $ changes from $110(4)
\to 88^{+4}_{-6}$ MeV and $m_s(M_\phi) $ from $132(6) \to
90^{+5}_{-11}$ MeV!  Second, extrapolations in sea quark mass from
``heavy'' dynamical $u$ and $d$ quarks, $i.e.$ in the range $2m_s 
\rightarrow m_s/2$, yield already (within the statistical uncertainty) 
very good agreement between $m_s(M_K)$ and $m_s(M_\phi)$. 
One needs to check, however, that this agreement persists
when the partially quenched analyses recommended
in~\cite{SRSpql00} are fully implemented. Third, the role of
the (so-far) neglected strange sea quarks remains to be
investigated. Note that so far all
lattice calculations indicate that the effect of adding dynamical
flavors is to lower the estimates of quark masses.  Fourth, the
renormalization constants used for all $n_f=2$ estimates are
perturbative; the corresponding non-perturbative calculations have yet
to be done.

In spite of these cautionary remarks, we consider the CP-PACS numbers
the current best lattice estimates.  We quote these as
$[m_u+m_d](2\ {\rm GeV}) = 6.8\pm 0.8\ {\rm MeV}$ and 
$m_s(2\ {\rm GeV})=89\pm 11\ {\rm MeV}$, where we
have doubled the error on $m_u+m_d$ quoted by
CP-PACS, based on that on $m_s$.

Finally, a brief comment on how CP-PACS numbers compare with
preliminary data reported at LATTICE 2000 (see review by
Lubicz~\cite{MqLUBICZ00}).  The numbers vary from $m_s = 90(5)$ MeV
(QCDSF-UKQCD~\cite{MqQCDSFUKQCD00}) to $m_s \approx 110$ MeV
(APE~\cite{MqLUBICZ00}, MILC~\cite{MqLUBICZ00}). We expect that within
the next couple of years a number of collaborations will have $n_f=2$
flavor results of numerical quality similar to those by CP-PACS. We are
therefore confident that a consistent picture of light quark masses
will emerge soon.

\pagebreak
\section{References}
\noindent



\ifx\href\undefined\def\href#1#2{{#2}}\fi
\def\spireshome{http://www.slac.stanford.edu/cgi-bin/spiface/find/hep/www?FORMAT=WWW&}
\def\xxxhome{http://xxx.lanl.gov/abs/}
{\catcode`\%=12
\xdef\spiresjournal#1#2#3{\noexpand\protect\noexpand\href{\spireshome
                          rawcmd=find+journal+#1%2C+#2%2C+#3}}
\xdef\spireseprint#1#2{\noexpand\protect\noexpand\href{\spireshome rawcmd=find+eprint+#1%2F#2}}
\xdef\spiresreport#1{\noexpand\protect\noexpand\href{\spireshome rawcmd=find+rept+#1}}
\xdef\spireskey#1{\noexpand\protect\noexpand\href{\spireshome key=#1}}
\xdef\xxxeprint#1{\noexpand\protect\noexpand\href{\xxxhome #1}}
}
\def\eprint#1#2{\xxxeprint{#1/#2}{#1/#2}}
\def\report#1{\spiresreport{#1}{#1}}
\def\nohref{}

\makeatletter
\def\putpaper{\@ifnextchar [{\@putpaper}{\@putpaper[]}}
\def\@putpaper[#1]{\edef\refpage{\the\count0}%
              \def\nohref{}%
              {\def\ {+}\def\nohref##1{}\edef\temp{\ifx\relax#1\relax
               \noexpand\spiresjournal{\journalname}{\volume}{\refpage}%
               \else\noexpand\xxxeprint{#1}\fi}\expandafter}\temp
               {\sfcode`\.=1000{\journalname} \journalformat}\egroup}
\def\putpage{\@ifnextchar [{\@putpage}{\@putpage[]}}
\def\@putpage[#1]{\edef\refpage{\the\count0}%
              \def\nohref{}%
              {\def\ {+}\def\nohref##1{}\edef\temp{\ifx\relax#1\relax
	       \noexpand\spiresjournal{\journalname}{\volume}{\refpage}%
               \else\noexpand\xxxeprint{#1}\fi}\expandafter}\temp
              {\refpage}\egroup}
\def\dojournal#1#2 (#3){\def\journalname{#1}\def\volume{#2}%
                         \def\refyear{#3}\afterassignment\putpaper\bgroup
                         \count0=}
\def\morepage{\afterassignment\putpage\bgroup\count0=}
\def\supresslink{\def\spiresjournal##1##2##3{}}

\def\APNY#1{\dojournal{Ann.\ Phys.\ \nohref{(N.\ Y.)}}{#1}}
\def\CMP#1{\dojournal{Comm.\ Math.\ Phys.}{#1}}
\def\IJMPC#1{\dojournal{Int.\ J.\ Mod.\ Phys.}{C#1}}
\def\IJMPE#1{\dojournal{Int.\ J.\ Mod.\ Phys.}{E#1}}
\def\JAP#1{\dojournal{J.\ App.\ Phys.}{#1}}

\def\MPA#1{\dojournal{Mod.\ Phys.\ Lett.}{A#1}}
\def\MPLA#1{\dojournal{Mod.\ Phys.\ Lett.}{A#1}}
\def\NP#1{\dojournal{Nucl.\ Phys.}{B#1}}
\def\NPA#1{\dojournal{Nucl.\ Phys.}{A#1}}
\def\NPB#1{\dojournal{Nucl.\ Phys.}{B#1}}
\def\NPBPS#1{\dojournal{Nucl.\ Phys.\ \nohref(Proc.\ Suppl.\nohref)}{\nohref B#1}}
\def\NPAPS#1{\dojournal{Nucl.\ Phys.\ \nohref(Proc.\ Suppl.\nohref)}{\nohref A#1}}
\def\NC#1{\dojournal{Nuovo Cimento }{#1}}
\def\PRL#1{\dojournal{Phys.\ Rev.\ Lett.}{#1}}
\def\PR#1{\dojournal{Phys.\ Rev.}{#1}}
\def\PRep#1{\dojournal{Phys.\ Rep.}{#1}}
\def\PRB#1{\dojournal{Phys.\ Rev.}{B#1}}
\def\PRC#1{\dojournal{Phys.\ Rev.}{C#1}}
\def\PRD#1{\dojournal{Phys.\ Rev.}{D#1}}
\def\PRE#1{\dojournal{Phys.\ Rev.}{E#1}}
\def\PL#1{\dojournal{Phys.\ Lett.}{#1B}}
\def\PLA#1{\dojournal{Phys.\ Lett.}{#1A}}
\def\PLB#1{\dojournal{Phys.\ Lett.}{#1B}}
\def\RMP#1{\dojournal{Rev.\ Mod.\ Phys.}{#1}}
\def\PREP#1{\dojournal{Phys.\ Rep.}{#1}}
\def\ZEITC#1{\dojournal{Z.\ Phys.}{C#1}}
\def\ZPC#1{\dojournal{Z.\ Phys.}{C#1}}

\def\ie{{\sl i.e.}}
\def\etc{{\it etc.}}
\def\ibid{{\it ibid}}
\def\etal{{\it et al.}}


\let\super=^
\catcode`\^=13
\def^{\ifmmode\super\else\initialsep\fi}

\def\journalformat{{\bf \volume}, \refpage\ (\refyear)}
\def\initialsep{}

\end{document}